\documentclass[review]{elsarticle}

\usepackage{lineno,hyperref}
\usepackage[skip=1.5 pt,font=scriptsize]{caption}
\modulolinenumbers[5]

\journal{J. Colloid Interface Sci.}
\usepackage{graphicx}
\usepackage{color}
\usepackage{epstopdf}
\DeclareGraphicsExtensions{.png}

\title{Chemical Vapor Deposition Growth of Bilayer Graphene in Between Molybdenum Disulfide Sheets}

\author[mymainaddress,mysecondaryaddress]{Wojciech Kwieci{\'n}ski}
\cortext[mycorrespondingauthor]{Corresponding author. w.kwiecinski@utwente.nl (Wojciech Kwieci{\'n}ski) }
\address[mymainaddress]{Physics of Interfaces and Nanomaterials, MESA+ Institute for Nanotechnology, University of Twente,  P.O.Box 217, 7500AE Enschede, The Netherlands.}
\address[mysecondaryaddress]{Faculty of Chemistry, Lodz University of Technology, Zeromskiego 116, 90-924 Lodz, Poland}
\address[mythirdaddress]{Physics of Fluids and J.M. Burgers Centre for Fluid Mechanics, MESA+ Institute for Nanotechnology, University of Twente,  P.O.Box 217, 7500AE Enschede, The Netherlands.}

\author[mymainaddress]{Kai Sotthewes}
\author[mymainaddress]{Bene Poelsema}
\author[mymainaddress]{Harold J. W. Zandvliet}

\author[mymainaddress,mythirdaddress,mycorrespondingauthor]{Pantelis Bampoulis}
\cortext[mycorrespondingauthor]{Corresponding author. p.bampoulis@utwente.nl (Pantelis Bampoulis) }

\begin{document}

\begin{frontmatter}

\begin{abstract}

Direct growth of flat micrometer-sized bilayer graphene islands in between molybdenum disulfide sheets is achieved by chemical vapor deposition of ethylene at about 800 $^\circ$C. The temperature assisted decomposition of ethylene takes place mainly at molybdenum disulfide step edges. The carbon atoms intercalate at this high temperature, and during the deposition process, through defects of the molybdenum disulfide surface such as steps and wrinkles. Post growth atomic force microscopy images reveal that circular flat graphene islands have grown at a high yield. They consist of two graphene layers stacked on top of each other with a total thickness of 0.74 nm.  Our results demonstrate direct, simple and high yield growth of graphene/molybdenum disulfide heterostructures, which can be of high importance in future nanoelectronic and optoelectronic applications. \\
\\
This article may be downloaded for personal use only. Any other use requires prior permission of the author and Elsevier publications. The following article appeared in J. Colloid Interface Sci., 505 (2017) 776-782 and may be found at https://doi.org/10.1016/j.jcis.2017.06.076.
\\

\end{abstract}

\begin{keyword}
\texttt MoS$_2$\sep  graphene\sep  2D materials\sep  chemical vapor deposition\sep heterostructures
\end{keyword}

\end{frontmatter}

\section{Introduction}

Graphene (Gr), an $sp^2$ hybridized carbon layer, has received tremendous attention due to its exceptional electronic and mechanical properties \cite{Novoselov200510451, Novoselov2005197, Novoselov2004666}. The discovery of the unique properties of graphene has triggered the revival of interest in other two-dimensional (2D) crystals, such as hexagonal boron nitride (h-BN), and transition metal dichalcogenides like molybdenum disulfide (MoS$_2$) \cite{Xu20133766}. These 2D atomic crystals bear great promise for various applications \cite{Xu20133766, wang2012electronics, bampoulis2017defect}. An elegant way to exploit the intrinsic properties of 2D crystals is using combinations of alternating weakly bound layers of different 2D crystals (e.g., graphene, h-BN and MoS$_2$), thus forming the so-called van der Waals heterostructures\cite{Novoselov2016, Geim2013419, Haigh2012764, Ponomarenko2011, Georgiou2013100, Yu20143055, bampoulis2017graphene}.  

Van der Waals heterostructures have attracted a lot of attention due to their promise for future electronics, catalysis and battery industry\cite{Novoselov2016, Yuan2016, rostamnia2016efficient, doustkhah2016covalently}. Their key properties arise from the fact that in heterostructures each material maintains its intrinsic electronic structure owing to the weak, van der Waals, interactions between adjacent layers.  For instance, graphene grown or mechanically transferred on h-BN has been shown to preserve its pristine properties \cite{Xue2011282}. The mobility of the charge carriers of graphene grown on h-BN via the chemical vapor deposition (CVD) method reaches 20.000 - 30.000 $cm^2 V^{-1} s^{-1}$ at room temperature \cite{Tang2013, Tang2015}. 

Stacking graphene and MoS$_2$ is of potential interest in nanoelectronics and optoelectronics due to the tunable band gap of MoS$_2$ and its indirect to direct band gap transition\cite{Wang2012699, Jin2013, Kam1982463, mak2010atomically, radisavljevic2011single, splendiani2010emerging} and the excellent electronic properties of graphene \cite{Novoselov200510451, Novoselov2005197, Novoselov2004666}. Many devices based on MoS$_2$/graphene heterostructures have already been fabricated. Field-effect transistors with a few-layers thick MoS$_2$ channel and graphene as a source, drain and gate electrodes have been realized and show an on/off ratio of $\sim$10$^6$. The gate electrode was isolated by a few layers of h-BN, creating a transistor entirely out of 2D material components \cite{all2dtransistor}. Furthermore, single-layer MoS$_2$ has been used as a tunneling barrier in a vertical graphene/MoS$_2$/graphene heterostructure creating a field-effect transistor with on/off ratios up to 10$^5$ at room temperature \cite{mos2verticalheterostructure}. The hole-current flowing through this device was almost entirely spin-polarized. Vertical heterostructures can be used as highly efficient photodetector and photocurrent generators, tunable with a back-gate \cite{Britnell947}. In addition, the mechanical properties of 2D materials may enable the fabrication of flexible electronic devices and help to scale-down electronics in the vertical dimension.

Therefore, the fabrication of high quality and large-area heterostructures is of fundamental and technological interest. At this moment, van der Waals heterostructures are prepared by PMMA (poly(methyl methacrylate)) assisted transfer of a 2D crystal on top of another. The 2D crystals have to be prepared in advance either by mechanical exfoliation, chemical vapor deposition or liquid phase exfoliation \cite{Novoselov2004666, Hernandez2008563, Li20091312}. The process is extremely complicated, challenging and it has a very low yield. It gives rise to structural complexities and uncertainties. For example, the mutual azimuthal orientation of the stacked layers is not well controlled and in most cases it is even completely random. Furthermore, contamination trapped between the interfaces during the transfer process can lead to charge inhomogeneities \cite{Bampoulis2015}. It is highly desirable to develop a reliable, high yield and simple method for the fabrication of van der Waals heterostructures. Recently, graphene has been epitaxially grown on h-BN, ZnO and ZnS substrates via templating chemical vapor deposition \cite{Tang2013, Tang2015, Yang2013792, Tang2012329, Son20113089, Liu20112032}. In addition, MoS$_2$/graphene has been grown by CVD \cite{Yu20143055}, by a hydrothermal method\cite{Koroteev201121199} and by an in-situ catalytic process by heating a Mo-oleate complex coated on sodium sulfate particles \cite{Fu2014}. 

Here we present the successful growth of micrometer-sized bilayer graphene in between MoS$_2$ sheets. The growth is achieved by decomposition of ethylene molecules at the highly reactive step edges of MoS$_2$. The resulting carbon atoms (or dimers) intercalate, through step edges or defects on the surface, between subsurface MoS$_{2}$ trilayers, where the nucleation and growth of circular graphene bilayers occur. Our work paves the way for the direct engineering of graphene/MoS$_2$ heterostructures suitable for fundamental research as well as device application.

\section{Experimental Methods}

Freshly cleaved MoS$_2$ samples ($5 \times10 mm^2$) were mechanically fixed on a Si substrate. The MoS$_2$/Si assembly was then mounted on a sample holder. The samples were subsequently inserted into a high vacuum chamber with a base pressure of $4 \times 10^{-8}$ mbar. Ethylene was used as the carbon precursor and its pressure during deposition was set to $1.1\times10^{-5}$ mbar. The underlying Si substrate was used to anneal the MoS$_2$ substrate by direct current heating. The temperature of the Si substrate was set to $800 \pm 20^\circ$C and was calibrated using a pyrometer. The duration of the process varied between different experiments. After the deposition, the samples stayed in vacuum for at least 12 hours during the cool down. The cooling was realized by switching off the current heating and letting the system approach room temperature.  Atomic Force Microscopy (AFM) and Scanning Tunneling Microscopy (STM) measurements were subsequently performed in N$_2$ environment by continuously purging with N$_2$ gas. The samples were imaged using tapping mode atomic force microscopy with an Agilent 5100 (Agilent) and SSS-FMR probes (Nanosensors) with a nominal spring constant of  2.8 N/m and a resonance frequency of 75 kHz. Scanning Tunneling Microscopy (STM) and Spectroscopy (STS) were performed in N$_2$ environment with an UHV variable temperature AFM/STM (Beetle TM, RHK Technology), with chemically etched W tips. For the X-ray Photoelectron Spectroscopy (XPS) measurements a Quantera SXM (Physical Electronics) was used. The X-rays were Al Kα, monochromatic at 1486.6 eV with a beam size of 200 $\mu m$. XPS measurements were done at several different locations on the samples. 

\section{Results}

\begin{figure}[h]
\centering
\includegraphics[width=0.5\textwidth]{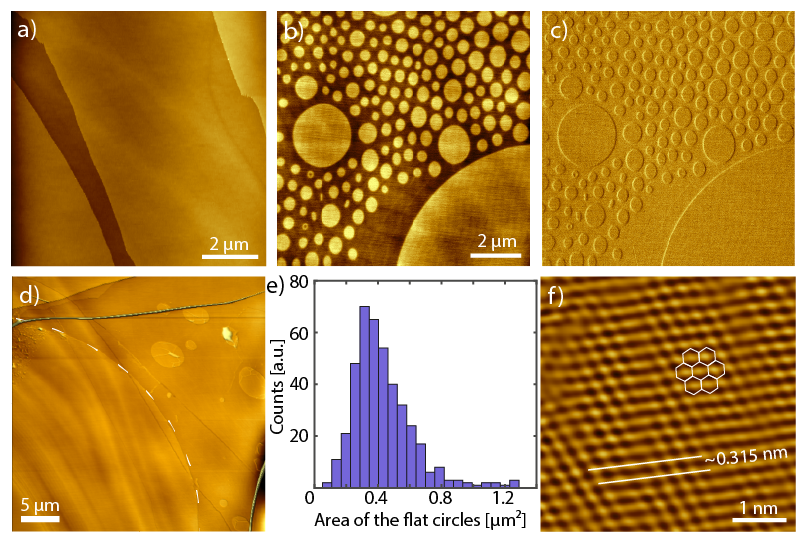}
\caption{(a) Tapping mode AFM topography image of a freshly cleaved MoS$_{2}$ surface. (b) AFM topography image and (c) the corresponding phase image of a MoS$_{2}$ surface after 5 minutes of CVD growth using ethylene. Circularly shaped flat layers with various radii are found intercalated between the MoS$_{2}$ tri-layers. The zero phase difference observed between the carbon-induced layers and the MoS$_{2}$ surface suggests that the carbon layers are intercalated between the MoS$_{2}$ tri-layers. (d) Topographic image of a MoS$_{2}$ surface with a large intercalated island, its contour is indicated by the white dashed circle segment. (e) Size distribution of the intercalated islands of region part of which is shown in panel (b).(f) Atomic-resolution LFM image of the MoS$_{2}$ surface on top of an intercalated island showing the lattice periodicity of the MoS$_{2}$ surface.}
\label{fig:Figure1}
\end{figure}

\begin{figure}[t]
\centering
\includegraphics[width=0.5\textwidth]{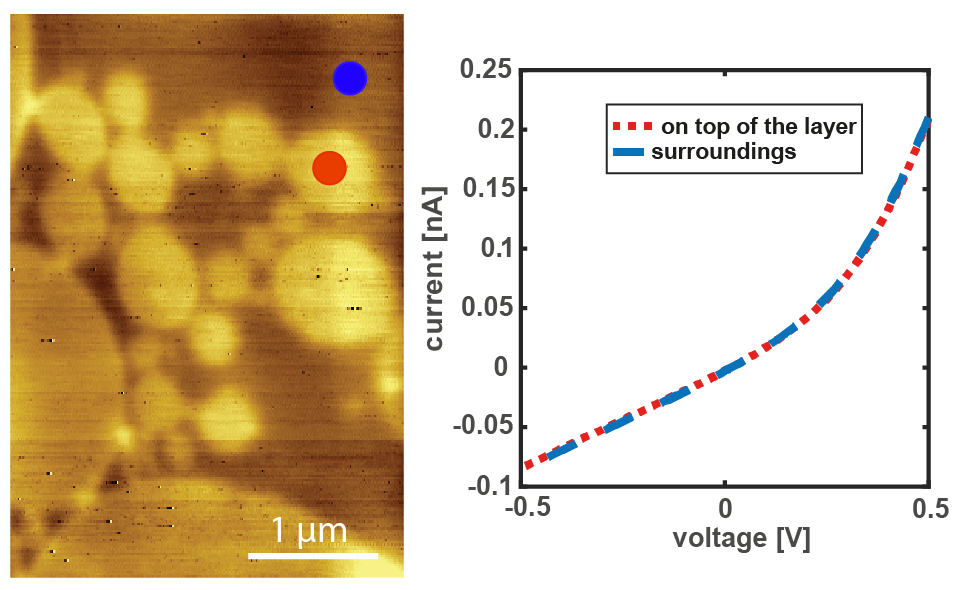}
\caption{STM topography image of carbon-induced islands intercalated between MoS$_{2}$ layers and Current-voltage (I(V)) spectroscopy curves recorded on top of an island and on the surrounding MoS$_{2}$ layer. The same characteristic I(V) behavior is measured, which unequivocally demonstrates that the island is intercalated and is covered with the same material as the surroundings, i.e., MoS$_{2}$. The I(V) curves appear to be metallic, since the set point used in order to record them is within the band gap of MoS$_2$. The set points are 0.2 nA, 0.5V}
\label{fig:Figure2}
\end{figure}

Figure \ref{fig:Figure1}a shows an AFM topographic image of a freshly cleaved MoS$_2$ surface. MoS$_2$ consists of stacked sulphur-molybdenum-sulphur tri-layers and the surface is smooth with a root mean square (RMS) roughness of about 45 pm and a low step density. After 5 min of CVD, parts of the surface are covered with circularly shaped flat layers with various radii, as shown in Figure \ref{fig:Figure1}b. Regions with a high concentration of small islands as well as regions with a low number density of large islands have been observed, as shown in Figures \ref{fig:Figure1}b and d. Their size distribution shows a large variation (see Figure \ref{fig:Figure1}e). These structures were formed only after the MoS$_{2}$ surface was heated to elevated temperatures and exposed to ethylene and thus will be referred to as carbon-induced layers. Phase images obtained simultaneously with AFM topographic images reveal that there is no phase difference between the layers and the intrinsic MoS$_2$ surface. Figure \ref{fig:Figure1}b shows an example of a topographic image and the corresponding phase image is given in Figure \ref{fig:Figure1}c. The absence of any phase contrast suggests that the carbon-induced structures are located underneath (instead of on top of) the outermost MoS$_2$ layer(s). In order to confirm this observation, we have performed high-resolution lateral force microscopy (LFM) on top of such a layer, we indeed measure the expected lattice periodicity of MoS$_2$, see Figure \ref{fig:Figure1}f. Supporting evidence was obtained by STS measurements. Current-Voltage Spectroscopy reveals that both the top of the layers and their surroundings display the same behavior, which corresponds to the bare MoS$_2$ electronic structure, see Figure \ref{fig:Figure2}. These findings clearly demonstrate that the carbon-induced layers are indeed located in between MoS$_2$ sheets.

\begin{figure}[t]
\centering
\includegraphics[width=0.6\textwidth]{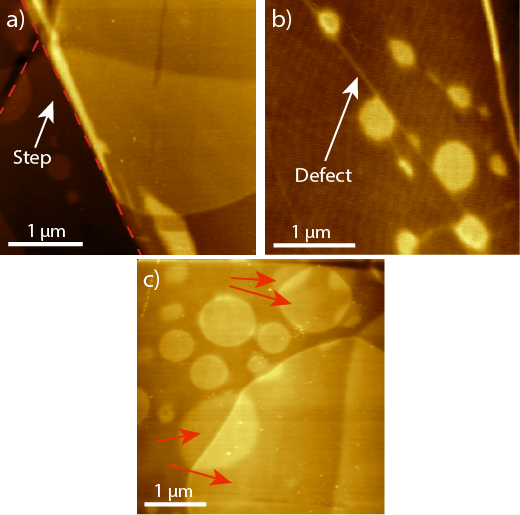}
\caption{(a) Tapping mode topography image showing a carbon-induced layer formed from a MoS$_{2}$ step or (b) defect, indicating that the intercalation of carbon happens through these pores. The red dashed lines in (a) indicate the location of the MoS$_2$ step edges. (c) Topography image of carbon-induced layers intercalated between different MoS$_{2}$ tri-layers. Parts of the carbon islands overlap rather than merging together. These islands grew between different MoS$_{2}$ tri-layers and therefore cannot coalesce. The red arrows indicate a few of the overlapping islands.}
\label{fig:Figure3}
\end{figure}

The carbon-induced layers are usually found in close vicinity to MoS$_2$ steps or defects, indicating that the intercalation of carbon takes place through these pores. An example is shown in Figure \ref{fig:Figure3}a and b, where carbon-induced layers were formed inward from a MoS$_2$ step and through an one-dimensional defect, respectively. These results clearly demonstrate the formation of carbon-induced layers between MoS$_2$ tri-layers. The MoS$_{2}$ surface is impermeable to small molecules such as ethylene and it is thus impossible for the ethylene molecules to penetrate through a perfect top MoS$_{2}$ tri-layer. Intercalation can occur between any of the MoS$_2$ sheets as long as there is a defect that allows intercalation of carbon atoms and seeds the growth process. Figure \ref{fig:Figure3}c shows an example of carbon induced islands intercalated between different MoS$_{2}$ tri-layers. Some of the islands overlap without merging which can only occur if the islands grow in different planes, i.e., between different MoS$_{2}$ tri-layers, which unequivocally demonstrates that the carbon layers are intercalated between MoS$_2$ trilayers.

\begin{figure}[!t]
\centering
\includegraphics[width=0.6\textwidth]{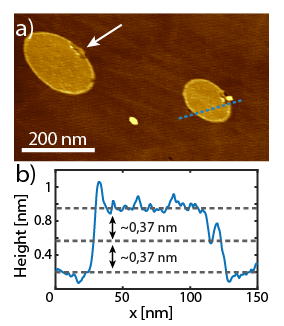}
\caption{(a) Tapping mode topography image of incomplete carbon-induced islands revealing that the islands are actually bilayers. The white arrows indicate the location of the defect on the top layer. (b) Line profile along the dashed line in (a) showing two levels with a height of approximately 0.37 nm.}
\label{fig:Figure4}
\end{figure}

Even though the vast majority of the intercalated islands do not have any defects, occasionally we observe incomplete islands (see Figure \ref{fig:Figure4}a). A line profile taken across such an island (Figure \ref{fig:Figure4}b) reveals that they actually consist of two levels with a height of $0.37 \pm 0.02$ nm per level, revealing that they are composed of bilayers. The total height of the intercalated bilayers is thus $0.74 \pm 0.04$ nm. The height values have been calibrated using the known height of a single tri-layer MoS$_2$ step, i.e., 0.615 nm. The height of the intercalated bilayers corresponds well to the observed height of graphene grown on h-BN\cite{Tang2015, Yang2013792}, which is slightly larger than the typical mono-atomic step height of A-B stacked graphite. The fact that bilayer islands dominate the AFM images suggest that they are by far thermodynamically more stable than single- or tri-layer islands.

\begin{figure}[h]
\centering
\includegraphics[width=0.6\textwidth]{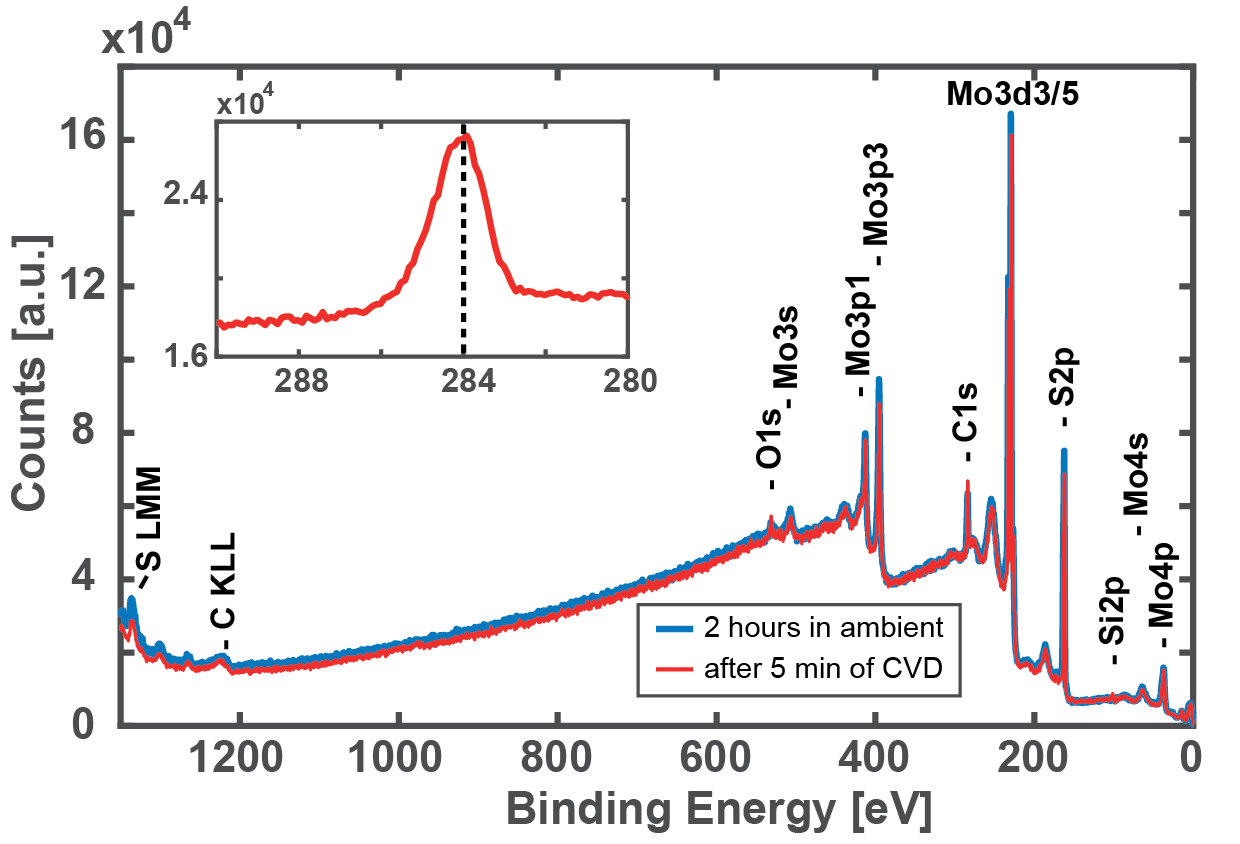}
\caption{Full range XPS spectrum of the MoS$_2$ sample after 5 minutes of CVD and the formation of carbon-induced layers. Inset: Carbon 1s peak from element specific measurement.}
\label{fig:XPSandAES}
\end{figure}

In an attempt to identify the composition of the carbon-induced bilayers we have performed XPS measurements, a characteristic spectrum is given in Figure \ref{fig:XPSandAES}. Five elements were found, namely: Mo (16-18\%), S (39-43\%), C (30-37\%), O (4.8-5.4\%) and Si (2.3-3.6\%). Mo and S signals originate from the MoS$_2$ substrate and both O and Si are probably the result of contamination induced during the transfer processes. The much higher carbon peak is naturally attributed to the carbon-induced structures or carbon contamination. Its binding energy was found to be between 284.0-284.5 eV. This value can be attributed either to pure carbon or carbohydrates \cite{bomben1992handbook}. It is not a value that is expected for carbides, which should be substantially smaller\cite{bomben1992handbook}. Smaller range element spectra done on two different areas showed that around 30-37 \% of all measured surface atoms were carbon atoms. The results do not give a clear answer on the exact composition of the circular carbon-induced layers since the carbon peaks can be a result of contamination by hydrocarbons on the MoS$_2$ surface during the sample transfer process.

Even though the XPS results do not provide a clear answer on the composition of the carbon-induced layers they clearly reveal no signs of carbides on the sample leaving as the only viable option that the structures are carbon bilayers. The fact that the height of these bilayers corresponds well to the height of a graphene layer on h-BN \cite{Yang2013792, Tang2015}, and it is very close to the step height of graphite, supports the idea that these carbon induced bilayers are actually graphene bilayers. This is further supported by the fact that graphite layers are by far the most thermodynamically stable carbon allotrope. In order to further validate the presence of graphene in our samples we did Raman mapping of locations with a very high density of graphene bilayers. Unfortunately, we could not find any detectable signal. The absence of a Raman signal is probably caused by the intercalated nature of the graphene bilayers under the multilayer MoS$_2$ cover, which would provide further support for our assessment. This can be potentially avoided by the use of thin MoS$_2$ samples, the thickness of which should not extend a few layers \cite{shi2012van}.

\begin{figure}[t]
\centering
\includegraphics[width=0.6\textwidth]{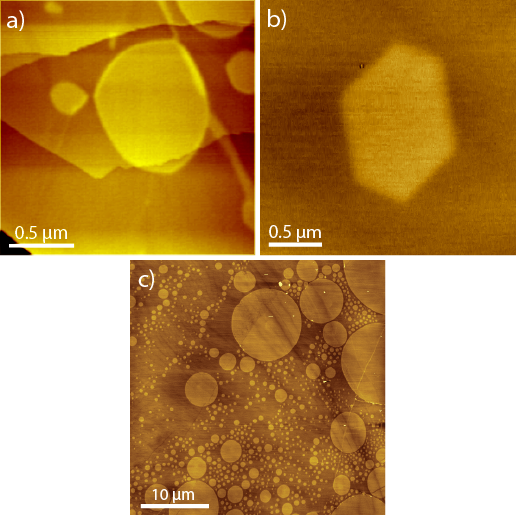}
\caption{(a-b) Topography images of intercalated layers  with faceted edges. (a) Occasionally, the circular layers have one or more faceted edges due to their local mono-crystalline structure. The structure in (b) is a mono-crystalline bilayer with perfect hexagonally faceted edges. The circularly shaped layers are probably polycrystalline and therefore the edges are less strongly faceted. (c) AFM topographic image of a large area, depicting several islands.}
\label{fig:Figure5}
\end{figure}

The circular shape of the graphene bilayers is an indication of polycrystallinity. Occasionally it can be seen that intercalated bilayers observed on the sample have some of their edges faceted, as shown in Figure \ref{fig:Figure5}a.  Graphene bilayers with hexagonal shapes are also occasionally found, see Figure \ref{fig:Figure5}b. The number density of hexagonal bilayers is significantly smaller than the circular ones, but they provide strong evidence of a hexagonal symmetry and thus crystallinity, in line with the graphene structure. Note that in conventional CVD growth of graphene islands on metal substrates, the interaction between carbon species and the metal substrate as well as temperature, define the morphology characteristics. Different island morphologies have been observed multiple times, ranging from fractal-like shapes to compact structures \cite{tetlow2014growth}. For instance, graphene flakes grown on Cu(111) have been shown to strongly depend on the substrate temperature and morphology during growth. Dendritic islands form at lower growth temperature and faceted compact islands at higher temperatures \cite{nie2011origin}. Circular flakes have been observed on graphene grown on copper and h-BN substrates \cite{Tang2013, wofford2010graphene}. In the latter case, the circular morphology was attributed also to the polycrystalline nature of the flakes \cite{Tang2013}. The weak van der Waals type interaction of the graphene bilayers with the MoS$_2$ walls could lead to the simultaneous growth of bilayers with different orientation with respect to MoS$_2$, growth and coalescence of these bilayers could give rise to a polycrystalline structure. Convolution induced by the MoS$_{2}$ cover will further smoothen the edges, an effect that becomes significant for thicker MoS$_{2}$ covers \cite{bampoulis2016coarsening}. 

\section{Discussion}

In this section, we attempt to rationalize the experimental observations and relate them to the growth mechanism. First, we emphasize that the subsurface high-temperature growth of islands makes it difficult (if not impossible) to experimentally access in situ information regarding their nucleation and growth mechanism. However, useful information that can shed some light on the underlying physics can be acquired from post-growth AFM images. We have established that the majority of the graphene islands are located near MoS$_{2}$ step edges and step bunches. This observation clearly shows the importance of these sites for the intercalation process. MoS$_{2}$ steps also provide a favorable place where temperature assisted decomposition of ethylene can occur. MoS$_2$ edges dominate the catalytic activity owing to their highly reactive nature \cite{chen2014ir, wang2014high, vang2008scanning, jaramillo2007}. The decomposition of ethylene therefore predominantly occurs at these locations. Most importantly, at these locations, molecular hydrogen can desorb from the surface. After the deposition and decomposition of ethylene at the steps, carbon (C) atoms/dimers can immediately intercalate between MoS$_2$ trilayers and diffuse `freely' owing to the weak van der Waals interactions between the trilayers, the lack of dangling bonds and the high temperature of the substrate. The high temperature during the deposition additionally warrants a high diffusion rate. The intercalation takes place through defects, step edges or wrinkles on the MoS$_2$ surface. The intercalation is probable, because the van der Waals forces connecting the MoS$_2$ tri-layers are relatively weak and the delamination energy that the atoms need to pay is extremely low \cite{Rydberg20031264021, Benavente200287}. A minority of the C atoms/dimers may also reside (temporarily) on top of the exposed MoS$_2$ layer.

\begin{figure*}[t]
\centering
\includegraphics[width=0.6\textwidth]{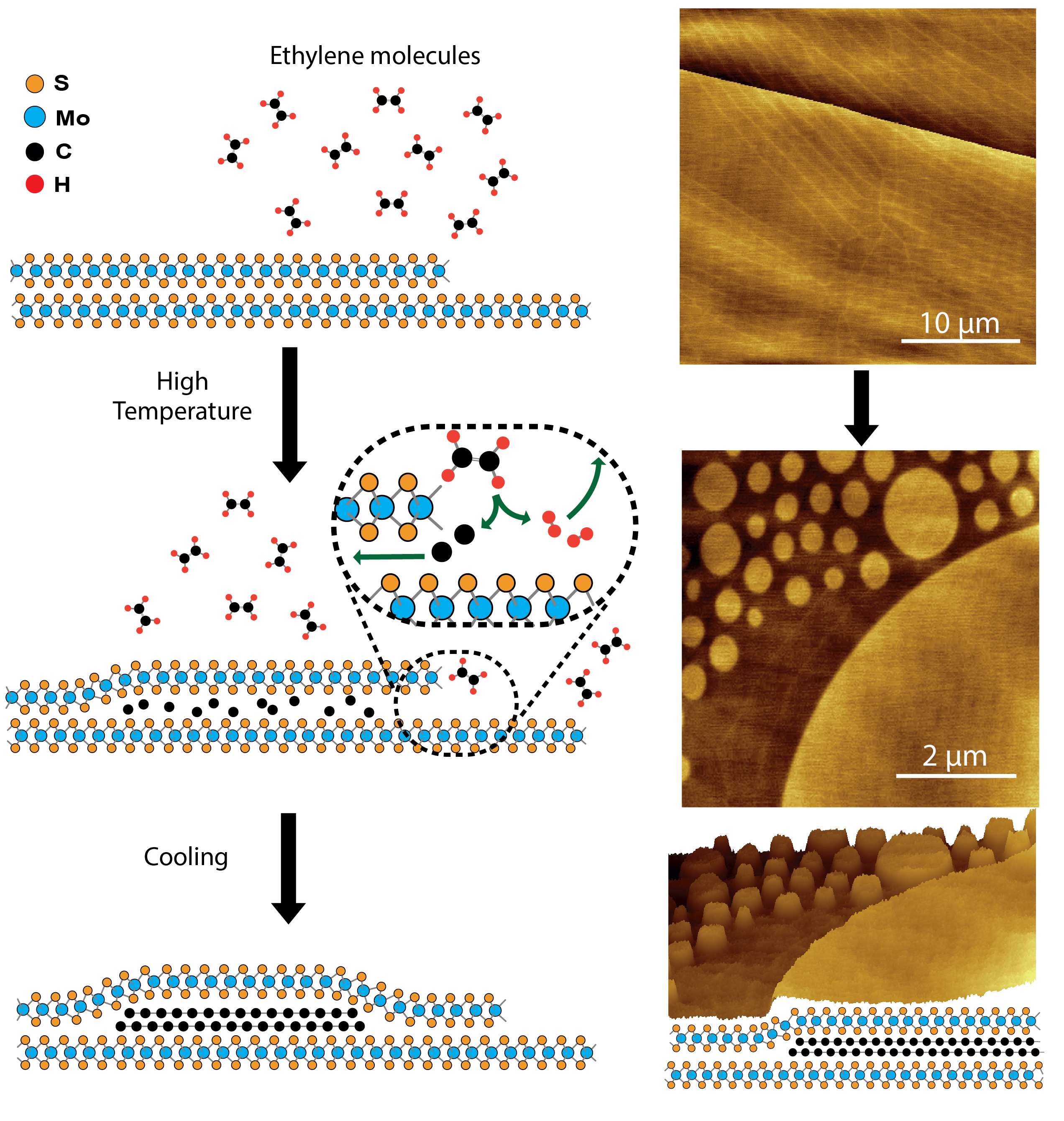}
\caption{Schematic drawing of the growth process of the intercalated graphene bilayers accompanied by the corresponding AFM images (right side). The ethylene molecules (black double spheres) break down predominantly at defects or step edges of the MoS$_{2}$ surface, where the metal is exposed and its catalytic action can favor the decomposition of ethylene (The yellow and blue spheres refer to sulfur and molybdenum atoms, respectively). Carbon atoms or dimers (black spheres) can freely diffuse randomly between MoS$_2$ trilayers because of the high surface temperature, the lack of dangling bonds of the MoS$_{2}$ and the weak van der Waals interactions. The carbon atoms/dimers intercalate between the MoS$_2$ tri-layers $via$ defects, step edges or wrinkles, where they eventually form graphene bilayers.}
\label{fig:Figure6}
\end{figure*}

\begin{figure}[h]
\centering
\includegraphics[width=0.6\textwidth]{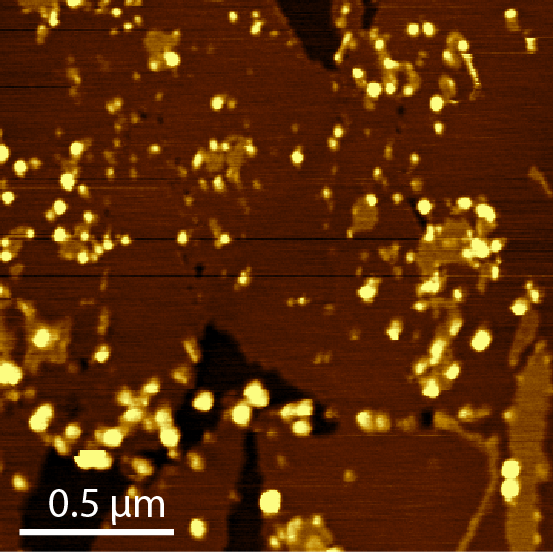}
\caption{AFM topographic image showing three dimensional particles formed on top of the MoS$_2$ after 10 mins of CVD.}
\label{fig:Figure7}
\end{figure}

The concentration of the carbon atoms increases as the deposition time increases. In a typical nucleation process, a graphene island will nucleate when a critical adatom concentration is reached \cite{Venables1984399}. However, as it is evident from several post growth AFM images the distribution of the islands is extremely inhomogeneous, see Figure \ref{fig:Figure5}c. In an attempt to rationalize the various observations we picture the following scenario. The highly mobile C species (atoms and probably dimers) form a gaseous mixture in between MoS$_2$ trilayers. These can travel large distances away from their origin (steps, defects etc.). The rate of this carbon diffusion depends on temperature and on the cross section for intercalation, i.e., the size of the defects. At the start of the deposition of ethylene the temperature is too high to form graphene (bi)layers since they are not stable. After terminating the deposition, the temperature is reduced continuously. At first nucleation occurs at regions with the highest local C concentration and these early condensates (graphene bilayers) can grow larger. Consequently the C concentration around these large features decreases, giving rise to the denuded zones surrounding them (Figs. \ref{fig:Figure5} and \ref{fig:Figure1}b). The continuous decrease of the temperature leads to reduced mobility and reaching of the continuously changing supersaturation conditions to form nuclei again, which remain now smaller because of the reduced C-mobility. Similar events are repeated upon further reaching of supersaturation conditions and concomitant nucleation. This course of events can be repeated several times, which can result in several subsequent stages of nucleation and therefore explain the strongly heterogeneous island sizes as obviously present in Fig. \ref{fig:Figure5}c. The exact resulting morphology will depend on a complex of factors such as cross section and nature of the defect and the exposed layer, deposition rate of ethylene, distance from the defect, rate of cooling, size of the MoS$_{2}$ trilayer package, etc. The process is summarized in the cartoon of Figure \ref{fig:Figure6}. A consequence of this scenario is that the size of the graphene circles can be influenced by controlling the cooling down trajectory. That should lead to much-enhanced homogeneity when the cooling rate is lowered. The fact that we mostly find intercalated islands suggests that they are thermodynamically far more favorable than islands on top of the surface. Occasionally and in particular at longer deposition times, clusters on top of the surface are formed. Their shape and structure is erratic, perhaps due to the available 3 degrees of freedom, as can be seen in Figure \ref{fig:Figure7}. Last but not least we have tested the stability of these structures at ambient conditions. We have stored a sample for about two months at ambient conditions. Remarkably, no visible changes were observed, the bilayers remained completely intact.

\section{Conclusions}

We have provided evidence for the growth of large (several micrometers in diameter) graphene bilayers between MoS$_2$ sheets by chemical vapor deposition of ethylene at 800$^\circ$C. The graphene bilayers have predominantly circular shapes and can grow up to several micrometers in diameter.  Their growth is realized by the intercalation of carbon atoms/dimers during the deposition process and through defects, such as step edges. The structure is facilitated by the decomposition of ethylene at reactive sites of MoS$_2$ such as Mo-terminated step edges. Structures on top of the MoS$_2$ surface have been also observed, but they are very scarce. This suggests that it is energetically favorable for carbon to intercalate in the MoS$_2$. Intercalation phenomena have been previously observed during the growth of group-IV semiconductors on MoS$_2$ \cite{zhang2016structural, yao2016growth}. However, in these studies the intercalated species did not form flat islands and the intercalation mechanism was not identified. Our approach demonstrates the direct growth of graphene/MoS$_2$ heterostructures and paves the way toward the realization of future 2D material based electronics. Understanding the nucleation and growth mechanisms of these layers is an important step toward optimization of the process that can lead to growth of large bilayers of graphene or even continuous films in between the MoS$_2$ sheets. Great enhancement of the homogeneity of the island sizes is expected from improved control of the temperature during the cooling down process. A further step should be taken in order to replace the bulk MoS$_2$ substrate with a few layers thick MoS$_2$ substrates which could enable precise control of the heterostructure thickness and is highly desirable for device applications. 
  
\section*{Acknowledgements}

The authors thank the Dutch Organization for Research (NWO, STW 11431) for financial support.

\section*{Bibliography}
\bibliography{Bibliography}
%\begin{tocentry}
%\centering
%\includegraphics[width=8 cm]{./TOC}
%\end{tocentry}
\end{document}